\documentclass[11pt]{article}
\usepackage{moriond}
\usepackage{graphicx}
\usepackage{amssymb}

\def\aa{{\em A\&A~}}
\def\aj{{\em AJ~}}
\def\apj{{\em ApJ~}}
\def\mnras{{\em MNRAS~}}
\def\nat{{\em Nature~}}
\def\na{{\em NewA~}}
\def\nar{{\em NewAR~}}

\def\cres{CR{\itshape e}s}
\def\crps{CR{\itshape p}s}
\begin{document}
\vspace*{4cm}
\title{Non-thermal emission from galaxy clusters:\\ a Pandora's vase for astrophysics}

\author{C. FERRARI}

\address{Universit\'e de Nice Sophia Antipolis, CNRS, Observatoire de la C\^ote d'Azur, \\ Laboratoire Cassiop\'ee, Nice, France}

\maketitle\abstracts{The existence of cosmic rays and weak magnetic
  fields in the intracluster volume has been well proven by deep radio
  observations of galaxy clusters. However a detailed physical
  characterization of the non-thermal component of large
  scale-structures, relevant for high-precision cosmology, is still
  missing. I will show the importance of combining numerical and
  theoretical works with cluster observations by a new-generation of
  radio, Gamma- and X-ray instruments.  }

\section{Introduction}

Deep radio observations of the sky have revealed the presence of
extended ($\sim$ 1 Mpc) radio sources in about 50 merging galaxy
clusters (see \cite{Feretti08,2008SSRv..134...93F} and references
therein). This diffuse radio emission is not related to unresolved
radio-galaxies, but rather to the presence of relativistic particles
($\gamma >>$1000) and magnetic fields of the order of $\mu$Gauss in
the intracluster volume. The physical mechanisms responsible for the
origin of this non-thermal intracluster component are matter of debate
(e.g. \cite{1977ApJ...212....1J,Govoni04}), as well as the effects of
intracluster cosmic rays (CRs) and magnetic fields on the
thermodynamical evolution and mass estimate of galaxy clusters
(e.g. \cite{2009arXiv0909.0270S,2010A&A...510A..76L}).  A deep
understanding of the evolutionary physics of {\em all} the different
cluster components (dark matter, galaxies, thermal and non-thermal
intracluster medium -- ICM) and of their mutual interactions is indeed
essential for high-precision cosmology with galaxy clusters
\cite{2005bmri.conf...77A}.

In the following, I will give an overview of our current knowledge of
the non-thermal component of galaxy clusters. I will also stress the
importance of a new generation of multi-wavelength telescopes -- such
as the {\it Low Frequency Array} ({\it LOFAR}), and the Gamma- and
hard X-ray (HXR) satellites {\it Fermi} and {\it NuSTAR} -- for a deep
understanding of the non-thermal cluster physics. The $\Lambda$CDM
model with H$_0$=70 km ${\rm s}^{-1} {\rm Mpc}^{-1}$, $\Omega_m=0.3$
and $\Omega_{\Lambda}=0.7$ has been adopted.

\section{The discovery of the non-thermal intracluster component:
  radio observations}\label{sec:disc}

The presence of intracluster CR electrons (\cres) and magnetic fields
was pointed out in 1970 by Willson \cite{1970MNRAS.151....1W}, whose
detailed radio analysis of the Coma cluster followed the first
detection in 1959 of a noticeably diffuse cluster radio source -- Coma
C -- by Large et al.  \cite{1959Natur.183.1663L}.

Diffuse cluster radio sources are very elusive. On the one hand their
low-surface brightness ($\sim\mu$Jy$/{\rm arcsec}^2$ at 1.4 GHz)
requires low angular resolution observations in order to achieve the
necessary signal-to-noise ratio. On the other hand complementary
high-resolution observations are needed in order to identify and
remove emission from point sources. Samples of clusters hosting
diffuse radio sources started to be available from the 90's
(e.g. \cite{1999NewA....4..141G}), with the advent of continuum radio
surveys such as the NVSS \cite{1998AJ....115.1693C}. It emerged that
the non-thermal plasma emitting at radio wavelengths could be not a
common property of galaxy clusters (see \cite{2002ASSL..272..197G} and
references therein). It was also found that a common feature of
intracluster radio sources is a steep synchrotron spectral slope
($\alpha\gtrsim$ 1~\footnote{S$(\nu) \propto \nu^{-\alpha}$},
\cite{2001ApJ...548..639K}). Based on the observed difference in other
physical properties (e.g. position in the host cluster, size and
morphology) a working classification with three main classes of
intracluster radio sources was soon adopted
\cite{1996IAUS..175..333F}:

\begin{figure*}
\centerline { \mbox{\includegraphics[width=160mm]{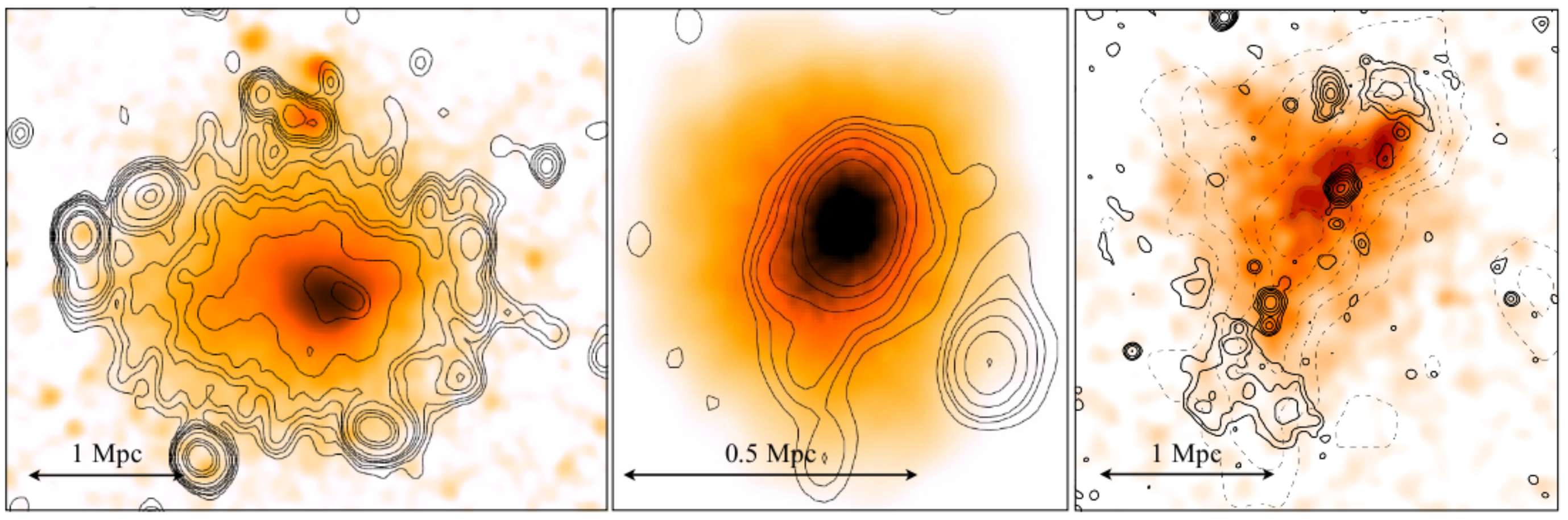}}}
\caption{The galaxy clusters Abell 2163 (left), RX J1347.5-1145
  (middle) and ZwCl 2341.1+0000 (right) observed in X-rays (brown
  scale image) and at radio wavelengths (solid contours) (adapted from
  \protect\cite{2001A&A...373..106F,2007A&A...470L..25G,2009A&A...506.1083V}). A2163
  is the hottest Abell cluster and it hosts one of the most luminous
  radio halos \protect\cite{2001A&A...373..106F}. A radio-mini halo is at the
  center of the most X-ray luminous cluster RX J1347.5-1145
  \protect\cite{2007A&A...470L..25G}. Double radio relics have been discovered
  in ZwCl 2341.1+0000 \protect\cite{2009A&A...506.1083V}. Diffuse radio
  emission has also been detected in this cluster along the optical
  filament of galaxies shown here by dashed contours
  \protect\cite{2010A&A...511L...5G}. \label{fig:hmhr}}
\end{figure*}

\begin{itemize}

\item {\it radio halos} are extended ($\gtrsim$1 Mpc) sources that have
  been detected at the centre of merging clusters. Their morphology is similar to
  the X-ray morphology of the cluster (Fig.~\ref{fig:hmhr}, left panel);

\item {\it radio mini-halos} are smaller sources ($\lesssim$500 kpc)
  located at the centre of cool-core clusters. They surround a
  powerful radio galaxy (Fig.~\ref{fig:hmhr}, middle panel);

\item {\it radio relics} have extensions similar to halos and are also
  detected in merging clusters, but they are usually located in the
  cluster outskirts and have an elongated morphology
  (Fig.~\ref{fig:hmhr}, right panel). In some clusters double relics
  have been detected (see
  \cite{2009A&A...494..429B,2009A&A...506.1083V} and references
  therein).

\end{itemize}

\noindent The discovery of a non-thermal intracluster component
through radio observations has opened a number of astrophysical
questions: How do cosmic rays and magnetic fields originate within the
intracluster volume? Are all the clusters hosting a non-thermal
component?  How does it affect the thermodynamical evolution and the
mass estimates of galaxy clusters? As detailed in the following
sections, new observational facilities will allow us to address most
of these open questions in the next few years.

\section{Non-thermal component of galaxy clusters: the known and unknown}

\subsection{Magnetic fields}\label{sec:b}

The intensity of intracluster magnetic fields can be measured
\cite{Carilli02,Govoni04}:

\begin{itemize}

\item through Faraday rotation measures (RM) of polarized radio sources
  within~/~behind clusters (current measurements: $\sim$1--10
  $\mu$Gauss);

\item by comparing synchrotron radiation from diffuse radio sources
  with non-thermal HXR emission due to Inverse Compton (IC) scattering
  of CMB photons by relativistic electrons (current measurements:
  $\sim$0.1--0.3 $\mu$Gauss);

\item by assuming energy equipartition between intracluster CRs and
  magnetic fields (current measurements: $\sim$0.1--1.0 $\mu$Gauss);

\item through the study of cold fronts in merging galaxy clusters
  (current measurements: $\sim$10 $\mu$Gauss).

\end{itemize}

\noindent The discrepancy between these different measurements can
indeed be related to the complex structure of intracluster magnetic
fields. Magnetic field models where both small and large scale
structures coexist must be considered, as recently shown by joint
observational and numerical studies
(e.g. \cite{2006A&A...460..425G,2008A&A...483..699G}). A radial
decline of the magnetic field strength has also been observed in
agreement with different magneto-hydrodynamic simulations
(\cite{2010A&A...513A..30B,2010arXiv1001.1058V} and references
therein). This can have important consequences in comparing, for
instance, volume averaged magnetic field measurements (such as those
obtained through the equipartition and IC methods) with RM estimates,
that are very sensitive to local variations in the magnetic field and
ICM structure.  Consistent magnetic field measurements have been
recently obtained in Coma by firstly determining a model of magnetic
field strength, radial profile and power spectrum, and then deriving
with the different methods an average magnetic field strength over the
same cluster volume \cite{2010A&A...513A..30B}. Finally, magnetic
field measures based on IC scattering of CMB photons have also to take
into account the controversial detection of HXR flux from galaxy
clusters (Sect. \ref{sec:multi}) and that radio ($\approx$1.4 GHz) and
XHR ($\approx$50 keV) radiations come from different populations of
intracluster relativistic electrons \cite{Carilli02}.

Magnetic fields at the observed intensity level ($\approx 1~\mu$Gauss)
could result from amplification of seed fields through adiabatic
compression, turbulence and shear flows associated to the hierarchical
structure formation process. Seed fields could have been created by
primordial processes and thus fill the entire volume of the universe,
or through different physical mechanisms, such as the ``Biermann
battery'' effect in merger and accretion shocks, or the outflow from
AGN and starburst galaxies in proto-clusters at $z\approx4-6$ (see
\cite{2008SSRv..134..311D} for a recent review). 

\subsection{Cosmic rays}\label{sec:CRs}

\begin{figure*}[h!]
\centerline { \mbox{\includegraphics[width=65mm]{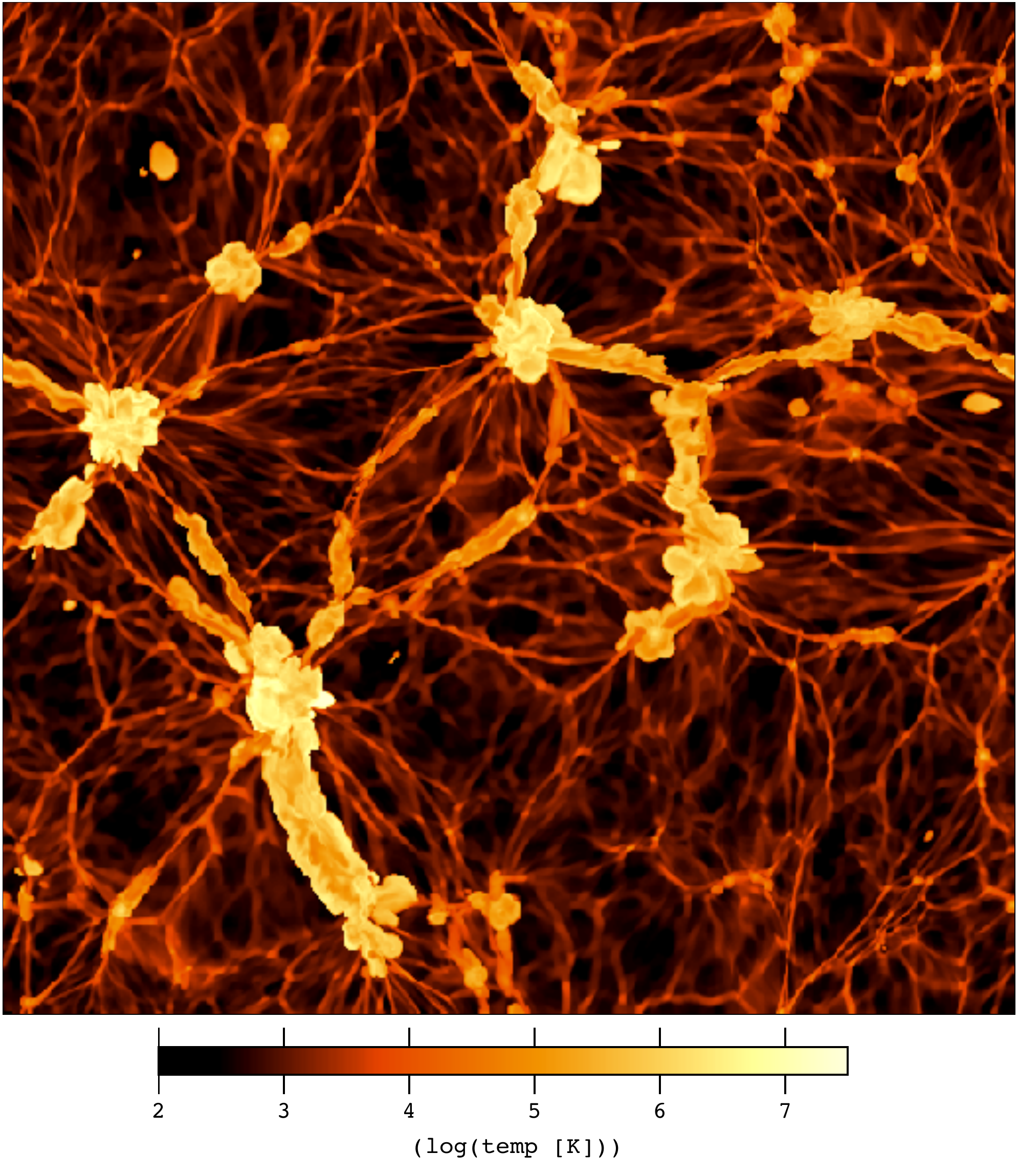} 
\hspace{2.0cm}
\includegraphics[width=65mm]{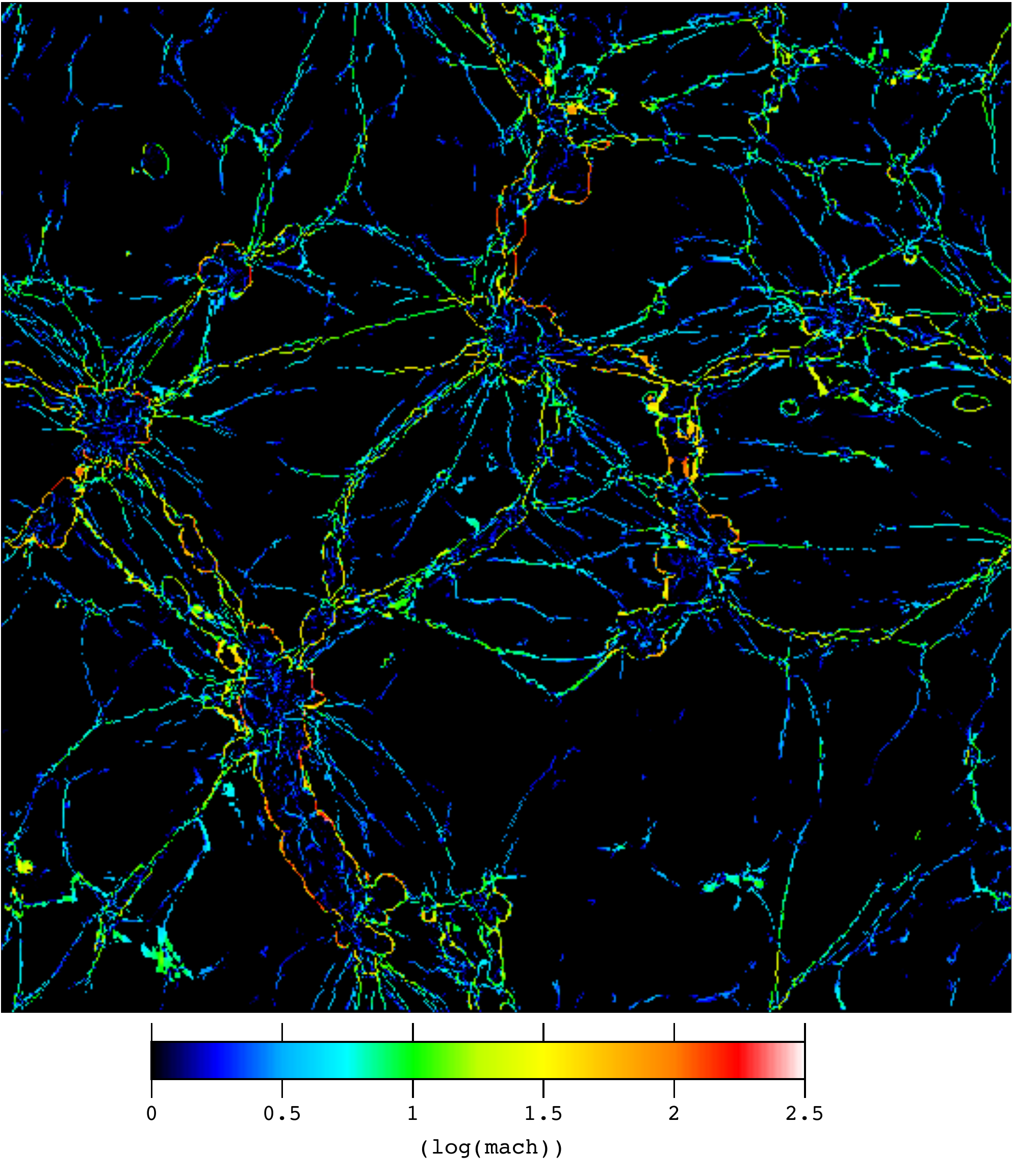} } }
\caption{Simulated maps of the gas temperature (left) and Mach number
  (right) in a region of side 80 Mpc (from
  \protect\cite{2009MNRAS.395.1333V}). Accretion shocks (due to large scale
  matter infall on massive clusters) have significantly higher Mach
  numbers compared to shocks that develop in the central regions of
  major merging clusters. Accretion shocks are expected to be highly
  efficient sources of particle acceleration, with steeper spectra of
  injection compared to the shocks internal to clusters. \label{fig:shock}}
\end{figure*}

Different mechanisms can produce CRs in galaxy clusters. Primary
relativistic particles can be accelerated by processes internal to
cluster galaxies, i.e. galactic winds or AGNs, and then ejected into
the intracluster volume. Intracluster CRs gyrate around magnetic field
lines which are frozen in the ICM. The expected diffusion velocity of
relativistic particles being of the order of the Alfv\'en speed
($\sim$ 100 km/s), CRs need $\gtrsim$10 Gyr to propagate over radio
halo and relic extensions. The radiative lifetime of relativistic
electrons is however much shorter ($\lesssim$0.1 Gyr) due to IC and
synchrotron energy losses. Therefore \cres~cannot simply be ejected by
active galaxies and propagate over the cluster volume, but they have
to be continuously (re-)accelerated {\it in situ}
\cite{2002ASSL..272....1S}. Electrons can be (re-)accelerated to GeV
energies by shocks and turbulence generated by major cluster mergers,
and to TeV energies at the strong accretion shocks
\cite{2000ApJ...542..608M}, where cold infalling material plunges in
the hot ICM of massive galaxy clusters and shock Mach numbers range
between 10 and a few 10$^3$ (see Fig.~\ref{fig:shock}).

The timescales for energy losses as well as the diffusion timescales
are instead longer than the Hubble time for CR protons (\crps). They
thus can be continuously accelerated both by internal and by external
processes, resulting in an effective accumulation of relativistic and
ultra-relativistic \crps~in clusters. Hadronic CRs can subsequently
produce Gamma-rays and secondary relativistic electrons through
inelastic collisions with the ions of the ICM (see
e.g. \cite{2009A&A...495...27A} and references therein). Other
possible physical mechanisms that could accelerate electrons up to
ultra-relativistic energies (TeV--PeV) are related to interactions of
CMB photons with ultra-relativistic \crps~ \cite{2005ApJ...628L...9I},
and/or very high energy intracluster Gamma-ray photons
\cite{2004A&A...417..391T}. Dark matter annihilation can also be a
source of secondary relativistic electrons and positrons
\cite{2010arXiv1001.4086P}.

Relativistic electrons observed at radio wavelengths can thus have a
secondary origin \cite{Dennison80}, and/or have been (re-)accelerated
by ICM shocks and turbulence developed during cluster mergers
\cite{Ensslin98,Brunetti01}. Current radio observational results are
mostly in agreement with this latter hypothesis (e.g.
\cite{2007ApJ...670L...5B}). The strongest point leading to this
conclusion is the fact that giant radio halos and relics have been
detected up to now only in merging clusters.  However, as detailed in
Sect.~\ref{sec:multi}, deeper radio, Gamma-ray and HXR observations
are required to get firm conclusions about the origin of intracluster
\cres.

At the light of current results different questions still need to be
answered. First of all we have to understand if relativistic electrons
are really hosted {\it only} in merging systems (as present radio
observations suggest) or if {\it all} clusters have a radio
halo/relic. In addition extended radio sources have not been detected
in {\it all} merging clusters. If shocks and turbulence related to
cluster mergers are the mechanisms responsible for electron
re-acceleration, the absence of observational evidence of intracluster
\cres~in several merging systems could be related to other physical
effects. The observed correlation between radio power and cluster mass
seems to indicate that only very massive cluster mergers are energetic
enough to accelerate electrons at relativistic velocities in the
intracluster volume \cite{2002ASSL..272..197G}. This scenario needs
however to be tested through higher sensitivity radio observations,
since the non-detection of radio halos/relics in many merging clusters
could be related to a lack of sensitivity of current instruments. Deep
future radio surveys (Sect. \ref{sec:persp}) will allow us to study
the evolution of the luminosity function of radio halos, giving
important constraints on current models for electron acceleration in
galaxy clusters \cite{Cassano06}.

\subsection{A multi-wavelength view of the non-thermal intracluster component} \label{sec:multi}

\begin{table}[t]
  \caption{The non-thermal ``Pandora's vase'' for galaxy clusters. We can expect multi-wavelength emission and particle acceleration from different kinds of interactions between: {\bf (first row)} intracluster magnetic fields, CMB photons and ICM ions, and {\bf (first column)} relativistic~/~ultra-relativistic cosmic rays accelerated by different possible physical mechanisms in galaxy clusters (see Sect. \ref{sec:CRs}). Note in addition that the interaction between CMB photons and intracluster Gamma-ray photons can produce ultra-relativistic \cres.\label{tab:pandora}}
  \vspace{0.2cm}
\begin{center}
\begin{tabular}{|c||c|c|c|}
\hline
 & & & \\
{\sc  \&} & {\scshape Magnetic Fields} & {\scshape CMB photons} & {\scshape ICM ions} \\
 & & & \\
\hline
\hline
 & & & \\
{\scshape Rel. \cres} & Radio emission & Hard X-rays &  \\
 & \footnotesize{(Synchrotron)} & \footnotesize{(Inverse Compton)} & \\
\hline
 & & & \\
{\scshape Rel. \crps} & & & Gamma-rays + Secondary CRe   \\
 & & & \footnotesize{(Hadronic collisions)}\\
\hline
 & & & \\
{\scshape Ultra-rel. \cres} & Hard X-rays & Gamma-rays &  \\
 & \footnotesize{(Synchrotron)}  & \footnotesize{(Inverse Compton)}  & \\
\hline
 & & & \\
{\scshape Ultra-rel. \crps} & & Ultra-rel. \cres &  \\
 &  & \footnotesize{(Bethe-Heitler)}  & \\
\hline
\end{tabular}
\end{center}
\end{table}

An increasing number of theoretical and numerical analyses
(e.g. \cite{2008MNRAS.385.1211P,2009RMxAC..36..201B}) are exploring
the possibility that a combination of CR protons and electrons of
primary and secondary origin can reproduce the multi-wavelength
radiation of the non-thermal intracluster component. Besides
synchrotron radio emission from GeV electrons and intracluster
magnetic fields, we can expect (see also Table~\ref{tab:pandora}):

\begin{itemize}

\item {\it HXR emission} from IC scattering of CMB photons by GeV
  electrons or from synchrotron emission of TeV electrons;

\item {\it Gamma-ray emission} from IC scattering of CMB photons by TeV
  electrons or from inelastic collision of \crps~with the ions of the CMB.

\end{itemize}

\noindent Radio synchrotron emission from galaxy clusters is now
firmly confirmed (Sect.~\ref{sec:disc}). Evidence of non-thermal (IC)
HXR emission from several clusters hosting diffuse radio sources has
been obtained mostly through the X-ray satellites {\it Beppo-SAX} and
{\it RXTE} (e.g. \cite{1999ApJ...513L..21F,1999ApJ...511L..21R}). The
detection and nature (thermal or non-thermal) of the HXR excess in
galaxy clusters is however strongly debated (\cite{Ferrari09} and
references therein). Up to now, only upper-limits have been derived
for the Gamma-ray emission of galaxy clusters, which imply a CR energy
density less than 5-20\% of the thermal cluster energy density. If we
assume intracluster magnetic fields of the order of $\mu$Gauss
(Sect.~\ref{sec:b}) and cluster radii of a few Mpc, it can easily be
derived that intracluster CR and magnetic field energy densities are
not far from equipartition \cite{2009arXiv0910.5715V}.

\section{Perspectives}\label{sec:persp}

In order to make a proper comparison between observational results and
current theoretical models about the origin and physical properties of
the non-thermal intracluster component, we need multi-frequency
observations of {\it statistical} samples of clusters hosting diffuse
radio sources. The study of galaxy cluster SED~\footnote{Spectral
  Energy Distribution} from Gamma-rays to low radio frequencies, for
instance, is essential to discriminate between the different particle
acceleration scenarios and to improve the measure of magnetic field
intensity (see e.g. \cite{2009A&A...502..711C}).

In the next decades several radio facilities -- such as {\it LOFAR},
{\it LWA}, {\it ASKAP}, {\it MeerKAT} and, last but not least, {\it
  SKA} -- will allow to significantly improve our knowledge about the
radio emission of the non-thermal intracluster component
(e.g. \cite{2004NewAR..48.1137F}). We are now assisting to the opening
of a spectral window largely unexplored by previous radio telescopes
($\nu<200$~MHz) thanks to {\it LOFAR}. Due to the steep synchrotron
spectrum oh halos and relics, the detection of diffuse cluster radio
sources is favored at this low frequencies (see Fig. 7 in
\cite{2008SSRv..134...93F}). The planned {\it LOFAR} All-Sky survey is
expected to detect about 350 radio halos at redshift $z\lesssim$0.6
\cite{2010A&A...509A..68C}.

\begin{figure*}[t]
\centerline { \mbox{\includegraphics[width=75mm]{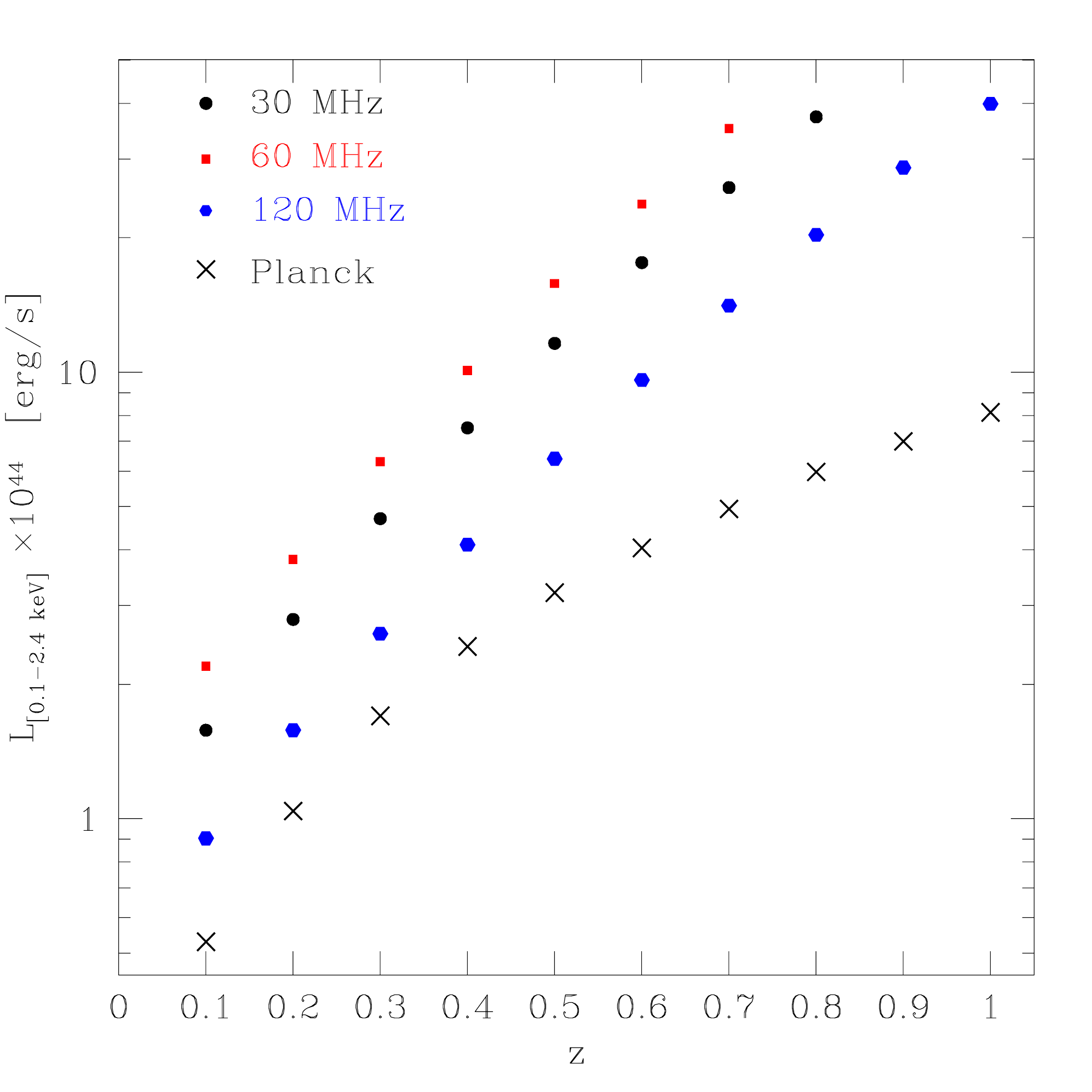} 
\hspace{1.0cm}
\includegraphics[width=75mm]{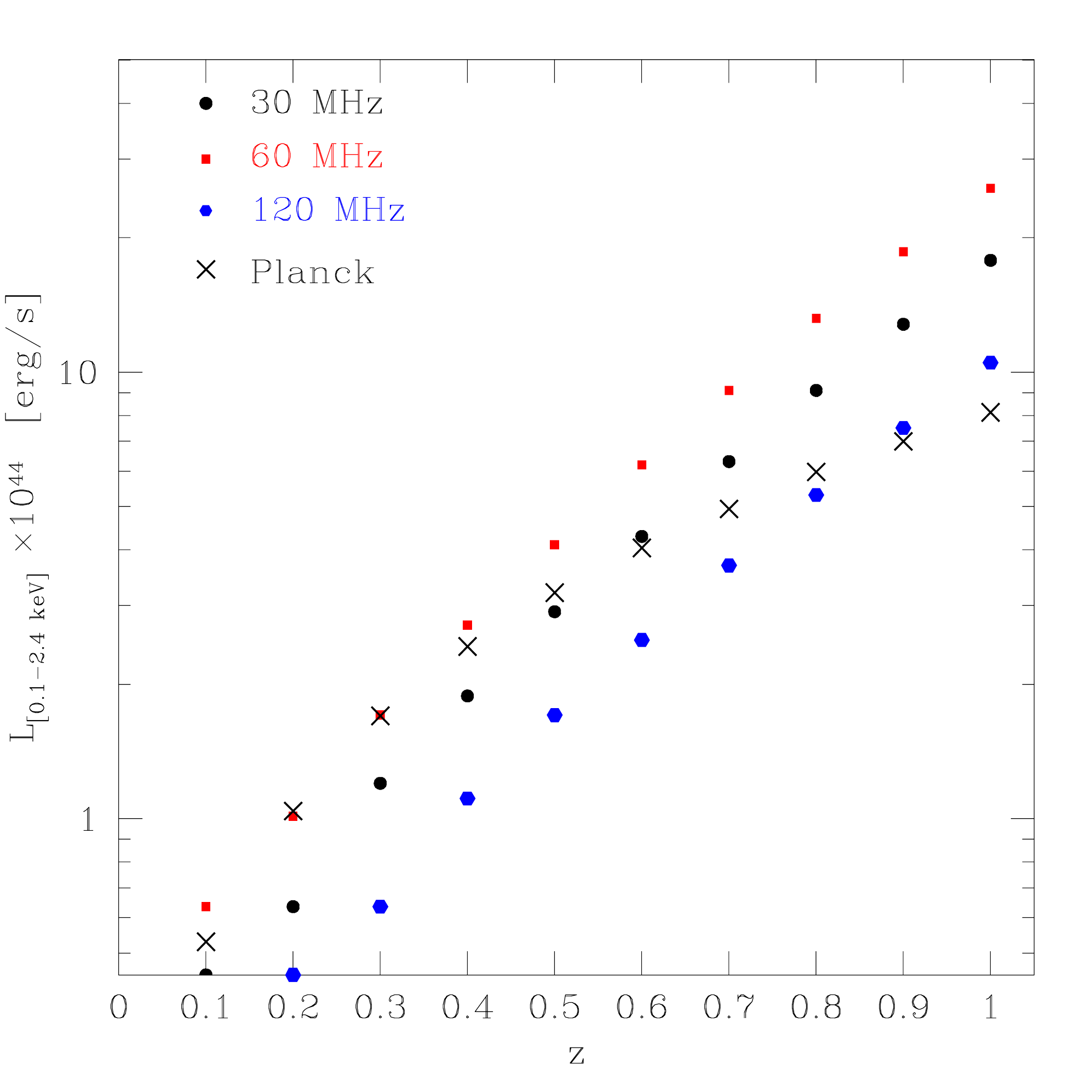} } }
\caption{Evolution with redshift of the X-ray luminosity limit of
  clusters whose diffuse radio emission can be detected with {\it
    LOFAR} at 30, 60 and 120 MHz down to the sensitivity limit of the
  All-Sky Survey (left) and Deep Survey (right), at a resolution of 21
  arcsec, assuming to detect at least 50\% of the radio flux at
  10$\sigma$ level (our estimates). The detection limits expected for
  the {\it Planck} cluster catalogue are shown with black crosses (courtesy
  of A. Chamballu and J. Bartlett).  \label{fig:LP}}

\end{figure*}

At other wavelengths and based on what detailed in previous sections,
important constraints about the non-thermal cluster emission are
expected from the Gamma-ray {\it Fermi} satellite, and from telescopes
observing in the HXR band, such as {\it NuSTAR} and, possibly, {\it
  IXO} (e.g. \cite{2009RMxAC..36..201B}). The detection of statistical
samples of radio halos and relics through on-going and up-coming radio
surveys (e.g. ``K.P. Extragalactic Surveys'' of {\it LOFAR}
\cite{2006astro.ph.10596R}, ``EMU'' survey of the {\it ASKAP}
telescope~\footnote{http://www.atnf.csiro.au/people/rnorris/emu/})
will need complementary multi-frequency projects for:

\begin{itemize}

\item obtaining complementary cluster catalogs to verify the presence
  of galaxy clusters corresponding to diffuse radio sources (see for
  instance the nice complementarity between the clusters that could be
  detected with {\it LOFAR} All-Sky and Deep Surveys and with {\it
    Planck}, Fig.~\ref{fig:LP});

\item getting a precise physical characterization of the detected
  cluster -- and in particular of its redshift, mass and dynamical
  state -- in order to test current models of CR acceleration.

\end{itemize}

\noindent To conclude, after the huge progress in the last fifteen
years of our knowledge of the evolutionary physics of cluster galaxies
and of the thermal ICM, we are now living in the ``golden age'' for
non-thermal cluster studies: the opening of the few spectral windows
largely unexplored by astronomical observations (i.e. the HXR,
Gamma-ray and low-frequency radio bands) will allow us to study the
non-thermal physics of galaxy clusters with unprecedented statistics
and thoroughness.

\section*{Acknowledgments}
I warmly thank the organizers of the Cosmology Moriond meeting for
having invited me to this interesting and enjoyable conference. Thanks
also go to Franco Vazza for his careful reading of the paper and his
very useful comments. I acknowledge financial support by the Agence
Nationale de la Recherche through grant ANR-09-JCJC-0001-01.


\section*{References}

\begin{thebibliography}{99}

\bibitem{Feretti08} Feretti, L. \& Giovannini, G., in {\em
    Lecture Notes in Physics}, Vol. 740, 143, A Pan-Chromatic View of
  Clusters of Galaxies and the Large-Scale Structure, ed. M.~Plionis,
  O.~L{\'o}pez-Cruz, \& D.~Hughes (Berlin Springer Verlag, 2008)

\bibitem{2008SSRv..134...93F} Ferrari, C. {\it et al.}, {\em
    Space Science Reviews} {\bf 134}, 93 (2008)

\bibitem{1977ApJ...212....1J} Jaffe, W.~J., \apj {\bf 212}, 1 (1977)

\bibitem{Govoni04} Govoni, F. \& Feretti, L, {\em
    IJMPD} {\bf 13}, 1549 (2004)

\bibitem{2009arXiv0909.0270S}
Sharma, P.  {\it et~al.}, {\em arXiv:0909.0270} (2009)

\bibitem{2010A&A...510A..76L}
Lagan\`a, T.~F. {\it et~al.}, \aa {\bf 510}, 76 (2010)

\bibitem{2005bmri.conf...77A} Arnaud, M., in {\em Background Microwave
    Radiation and Intracluster Cosmology}, 77, Proceedings of the
  International School of Physics ''Enrico Fermi'', ed. F.~Melchiorri
  \& Y.~Rephaeli (IOS Press, The Netherlands, and Societ\`a
  Italiana di Fisica, Bologna, Italy, 2005)

\bibitem{1970MNRAS.151....1W} Willson, M.~A.~G., \mnras{\bf 151}, 1, (1970)

\bibitem{1959Natur.183.1663L}
Large, M.~I. {\it et~al.}, \nat {\bf 183}, 1663 (1959)

\bibitem{1999NewA....4..141G}
Giovannini, G.  {\it et~al.}, \na {\bf 4}, 141 (1999)

\bibitem{1998AJ....115.1693C} Condon, J.~J. {\it et~al.}, \aj {\bf 115}, 1693 (1998)

\bibitem{2002ASSL..272..197G} Giovannini, G. \& Feretti, L., 2002, in {\em
  Astrophysics and Space Science Library}, Vol. 272, 197, Merging Processes
  in Galaxy Clusters, ed. L.~Feretti, I.~M.~Gioia, \& G.~Giovannini,
  (Kluwer Academic Publishers, Dordrecht, 2002)

\bibitem{2001ApJ...548..639K} Kempner, J.~C. \& Sarazin, C.~L.,
  \apj {\bf 548}, 639 (2001)

\bibitem{1996IAUS..175..333F} Feretti, L. \& Giovannini, G., in {\em
    IAU Symposium}, Vol. 175, 333, Extragalactic Radio Sources,
  ed. R.~D.~Ekers, C.~Fanti, \& L.~Padrielli (1996)

\bibitem{2001A&A...373..106F} Feretti, L. {\it et~al.},
  \aa {\bf 373}, 106 (2001)

\bibitem{2007A&A...470L..25G} Gitti, M.  {\it et~al.},
  \aa {\bf 470}, 25 (2007)

\bibitem{2009A&A...494..429B} Bonafede, A. {\it et~al.},
  \aa {\bf 494}, 429 (2009)

\bibitem{2009A&A...506.1083V} van Weeren, R. J.  {\it et~al.},
  \aa {\bf 506}, 1083 (2009)

\bibitem{2010A&A...511L...5G} Giovannini, G. {\it et~al.},
  \aa {\bf 511}, 5 (2010)

\bibitem{Carilli02} Carilli, C.~L. \& Taylor, G.~B., {\em
    ARA\&A} {\bf 40}, 319 (2002)

\bibitem{2006A&A...460..425G} Govoni, F.   {\it et al.},
  \aa {\bf 460}, 425 (2006)

\bibitem{2008A&A...483..699G} Guidetti, D.  {\it et al.},
  \aa {\bf 483}, 699 (2008)

\bibitem{2010A&A...513A..30B} Bonafede, A. {\it et al.},
  \aa {\bf 513}, 30 (2010)

\bibitem{2010arXiv1001.1058V} Vacca, V.  {\it et al.}, {\em
    arXiv:1001.1058} (2010)

\bibitem{2008SSRv..134..311D} Dolag, K., {\it et al.}, {\em
    Space Science Reviews} {\bf 134}, 311 (2008)

\bibitem{2002ASSL..272....1S} Sarazin, C.~L.,  in {\em
  Astrophysics and Space Science Library}, Vol. 272, 1, Merging Processes
  in Galaxy Clusters, ed. L.~Feretti, I.~M.~Gioia, \& G.~Giovannini,
  (Kluwer Academic Publishers, Dordrecht, 2002)

\bibitem{2000ApJ...542..608M} Miniati, F. {\it et~al.},
  \apj {\bf 542}, 609 (2000)

\bibitem{2009A&A...495...27A} Aharonian, F.  \aa {\bf 495}, 27 (2009) 

\bibitem{2009MNRAS.395.1333V} Vazza, F.  {\it et~al.},
  \mnras {\bf 395}, 1333 (2009)

\bibitem{2005ApJ...628L...9I} Inoue, S.  {\it et~al.},
  \apj {\bf 628}, 9 (2005)

\bibitem{2004A&A...417..391T} Timokhin, A.~N. {\it et~al.},
  \aa {\bf 417}, 391 (2004)

\bibitem{2010arXiv1001.4086P} Profumo, S. \& Ullio, P., {\em
    arXiv:1001.4086} (2010)

\bibitem{Dennison80} Dennison, B., \apj {\bf 239}, L93 (1980)

\bibitem{Ensslin98} En{\ss}lin T.~A. {\it et~al.},
  \aa {\bf 332}, 395 (1998)

\bibitem{Brunetti01} Brunetti, G.  {\it et~al.},
  \mnras {\bf 320}, 365 (2001)

\bibitem{2007ApJ...670L...5B} Brunetti, G.  {\it et~al.},
  \aa {\bf 670}, 5 (2007)

\bibitem{Cassano06} Cassano, R. {\it et~al.},
  \mnras {\bf 369}, 1577 (2006)

\bibitem{2008MNRAS.385.1211P} Pfrommer, C. {\it et~al.},
  \mnras {\bf 385}, 1211 (2008)

\bibitem{2009RMxAC..36..201B}
Brunetti, G., in {\em Revista Mexicana de Astronomia y Astrofisica Conference
  Series}, Vol.~36, 201 (Revista Mexicana de Astronomia y Astrofisica Conference
  Series 2009)

\bibitem{1999ApJ...513L..21F} Fusco-Femiano, R. {\it et~al.},
  \apj {\bf 513}, L21 (1999)

\bibitem{1999ApJ...511L..21R} Rephaeli, Y.  {\it et~al.},
  \apj {\bf 511}, L21 (1999)

\bibitem{Ferrari09} Ferrari, C. {\em AIPC} {\bf 1126}, 277 (2009)

\bibitem{2009arXiv0910.5715V}
Vannoni, G. {\it et~al.}, {\em arXiv:0910.5715} (2009)

\bibitem{2009A&A...502..711C} Colafrancesco, S. \& Marchegiani, P.,
  \aa {\bf 502}, 711 (2009)

\bibitem{2004NewAR..48.1137F} Feretti, L.  {\it et~al.},
  \nar {\bf 48}, 1137 (2004)

\bibitem{2010A&A...509A..68C} Cassano, R. {\it et~al.},
  \aa {\bf 509}, 68 (2010)

\bibitem{2006astro.ph.10596R}
R\"ottgering, H.~J.~A.  {\em astro-ph/0610596} (2006)

\end{thebibliography}
\end{document}